\documentclass[useAMS,usenatbib,usedcolumn]{mnras}

\usepackage{graphics,epsf}
\usepackage{amsmath}                
\usepackage{amsfonts}              
\usepackage{amssymb}                
\usepackage{epsfig}
\usepackage{epstopdf}
\usepackage{multirow}
\usepackage[para,online,flushleft]{threeparttable}

\usepackage{xcolor}
\definecolor{redak}{rgb}{0.9,0.15,0.05}

\def \kms{~\rm{km~s^{-1}}}

\def \msyr{~\rm{M_{\odot}}~\rm{yr^{-1}}}
\def \cm{~\rm{cm}}
\def \s{~\rm{s}}
\def \km{~\rm{km}}

\def \K{~\rm{K}}
\def \g{~\rm{g}}

\def \AU{~\rm{au}}
\def \erg{~\rm{erg}}

\def \yr{~\rm{yr}}

\def \days{~\rm{days}}

\def \rmModot{~\rm{M_{\sun}}}

\title[SN impostors by jets from NS companion]{Common envelope jets supernova (CEJSN) impostors resulting from a neutron star companion}
\author[A. Gilkis, N. Soker \& A. Kashi]{Avishai Gilkis$^{1}$\thanks{Contact e-mail: \href{agilkis@ast.cam.ac.uk}{agilkis@ast.cam.ac.uk}}, Noam Soker$^{2,3}$\thanks{Contact e-mail: \href{soker@physics.technion.ac.il}{soker@physics.technion.ac.il}}, Amit Kashi$^{4}$\thanks{Contact e-mail: \href{kashi@ariel.ac.il}{kashi@ariel.ac.il}}
\\
$^{1}$ Institute of Astronomy, University of Cambridge, Madingley Road, Cambridge, CB3 0HA, UK\\
$^{2}$ Department of Physics, Technion -- Israel Institute of Technology, Haifa 3200003, Israel\\
$^{3}$ Guangdong Technion Israel Institute of Technology, Shantou, Guangdong Province 515069, China\\ 
$^{4}$ Department of Physics, Ariel University, Ariel, POB 3, 40700, Israel
}

\begin{document}

\pagerange{\pageref{firstpage}--\pageref{lastpage}} \pubyear{2018}

\maketitle
\label{firstpage}

\begin{abstract}
We propose a new type of repeating transient outburst initiated by a neutron star (NS) entering the envelope of an evolved massive star, accreting envelope material and subsequently launching jets which interact with their surroundings. This interaction is the result of either a rapid expansion of the primary star due to an instability in its core near the end of its nuclear evolution, or due to a dynamical process which rapidly brings the NS into the primary star. The ejecta can reach velocities of $\approx 10^4 \kms$ despite not being a supernova, and might explain such velocities in the 2011 outburst of the luminous blue variable progenitor of SN~2009ip. The typical transient duration and kinetic energy are weeks to months, and up to $\approx 10^{51} \erg$, respectively. The interaction of a NS with a giant envelope might be a phase in the evolution of the progenitors of most NS-NS binary systems that later undergo a merger event. If the NS spirals in all the way to the core of the primary star and brings about its complete disruption we term this a `common envelope jets supernova' (CEJSN), which is a possible explanation for the peculiar supernova iPTF14hls. For a limited interaction of the NS with the envelope we get a less luminous transient, which we term a CEJSN impostor.
\end{abstract}


\section{INTROCUTION}
\label{sec:intro}

Observations of core collapse supernovae (CCSNe) and their analysis indicate that a non-negligible fraction of their progenitors eject a substantial amount of mass tens of years to several days before explosion (e.g., \citealt{Foleyetal2007, Mauerhanetal2013, Ofeketal2013, SvirskiNakar2014, Moriya2015, Goranskijetal2016, Tartagliaetal2016, Arcavietal2017, Marguttietal2017, Nyholmetal2017, Reillyetal2017, Yaronetal2017, BoianGroh2018, Liuetal2018, Pastorelloetal2018}). Just before collapse, nuclear reactions in the core release a huge amount of energy. Most of it is carried away by neutrinos (e.g., \citealt{ZirakashviliPtuskin2016}), but some fraction of this energy might find its way to the stellar envelope.
  
Mechanisms that might carry energy from the violent nuclear burning to the envelope, like waves \citep{QuataertShiode2012, ShiodeQuataert2014} and magnetic activity \citep{SokerGilkis2017a}, are likely to cause mainly envelope expansion rather than mass ejection (e.g., \citealt{Soker2013, McleySoker2014, Fuller2017}). One way to utilize the expanding envelope for the ejection of mass is by a binary interaction. A stellar companion that was detached from the envelope of the primary star, i.e., the progenitor of the supernova, before the envelope was inflated starts to accrete mass from the inflated envelope. The mass flows onto the secondary star through an accretion disc, and this disc launches jets that remove mass from the inflated envelope (e.g., \citealt{KashiSoker2010a, Soker2013, McleySoker2014, DenieliSoker2018}). 

A similar type of outburst might take place in the case that the primary star suffers a rapid expansion in late stages of evolution even when it is yet far from explosion. The most prominent example is the Great Eruption of the binary system Eta Carinae (on the Great Eruption itself see, e.g., \citealt{DavidsonHumphreys2012}). In earlier papers two of us suggested that accretion of mass from the primary star onto the secondary more compact star powered the Great Eruption (e.g., \citealt{KashiSoker2010a}).

Part of the accretion energy is channeled to light, by the accretion process itself, by the collision of the jets with the envelope, and/or from the collision of the freshly ejected envelope mass with previously ejected slower mass. The bright event might mimic a supernova explosion, and hence it is referred to as a supernova impostor. Supernova impostors overlap with major eruptions of luminous blue variables (LBVs), and both groups are part of the larger and heterogeneous group of intermediate luminosity optical transients (ILOTs; \citealt{KashiSoker2016}).

If the accreting companion is a compact object, the energy release will be larger, and the conditions in the accretion flow might be extreme in terms of density and accretion rate. Several studies consider the possible interaction between a neutron star (NS) or a black hole (BH) and the envelope (or even the core) of its larger companion, as mechanisms for gamma-ray bursts \citep{FryerWoosley1998,ZhangFryer2001}, supernova-like explosions \citep{BarkovKomissarov2011,Chevalier2012,SokerGilkis2018}, or both \citep{Thoneetal2011,Fryeretal2014}. Jets might be launched from accretion onto a white dwarf (WD) or a main sequence (MS) star  \citep{Soker2004}, or, more pertinent for the current study, from an accretion disc forming around a NS \citep*{ArmitageLivio2000,Papishetal2015,SokerGilkis2018}. The case of a NS companion that by launching jets explodes and terminates the evolution of the primary giant star was termed by \cite{SokerGilkis2018} a common envelope jets supernova (CEJSN).

Two major uncertainties in the process where a compact star accretes mass inside the envelope of a giant star are the accretion rate and the formation of an accretion disc. Different studies, in particular hydrodynamic simulations of accretion onto a compact object inside a common envelope (e.g., \citealt*{RasioShapiro1991,Fryeretal1996,Lombardietal2006,RickerTaam2008,MacLeod2015,MacLeod2017}), have reached different conclusions on the accretion rate and on whether accretion discs are formed or not. There are two key processes that facilitate the formation of an accretion disc. Firstly, the jets themselves removes energy and high entropy material from the vicinity of the accreting object \citep*{Shiberetal2016} and by that reduce the pressure near the accreting object. \cite{Chamandyetal2018} show in their hydrodynamical simulations that this energy removal allows a high accretion rate. If this energy removal by the jets is not considered, then pressure is built-up near the accreting object and the accretion rate is much lower (e.g. \citealt{RickerTaam2012, MacLeodRamirezRuiz2015b}). Secondly, it is very likely that an accretion disc is formed before the compact companion enters the envelope and it continues to exist inside the envelope \citep{Staffetal2016}. We return to discuss these two processes in the relevant sections.

We note also that jets (and disc winds) remove angular momentum from the accretion flow. \cite{MacLeodRamirezRuiz2015b} argue that the steep density gradient in the envelope imposes an angular momentum barrier to accretion onto a compact object in the envelope. Yet during Roche-lobe overflow mass is flowing onto the compact object from one side, and an accretion disc does form. In this case the envelope of the donor is synchronized with the orbital motion. Even if there is no synchronization, rotation of the envelope, that must exist to some degree, facilitates the formation of an accretion disc. \cite{Staffetal2016} include envelope rotation and find the formation of an accretion disc before the companion enters the envelope. Once an accretion disc exists, jets can remove angular momentum and maintain the disc. Even if the initial disc is small, friction within the disc leads to its expansion and the material in the disc collides then with the accreted gas at increasing distances. \cite{Chamandyetal2018} do not include envelope rotation, but nonetheless find that an accretion disc can be formed inside the common envelope. They do remove mass and energy from the vicinity of the accreting body, as we expect jets to do. For the above reasons, in our study we assume that such accretion discs form and launch jets. We expect the formation of these discs to start while the companion is still outside the envelope of the giant, and this might be an important feature in a thoroughly self-consistent model.

In the present study we consider non-terminal eruptions that can be classified as supernova impostors, or more generally as ILOTS, by an accreting NS companion that launches jets while orbiting inside the envelope or while grazing the envelope. We term this a CEJSN impostor. In some of these cases, perhaps when the orbit is rather eccentric, the NS will exit the envelope after the eruption and the process can repeat itself. In section \ref{sec:scenarios} we discuss scenarios which can bring a NS into the envelope of a supergiant star, and initiate the accretion and outflow which powers an energetic outburst. In section \ref{sec:parameters} we derive scaled relations to show the outburst characteristics, focusing on the scenario of rapid expansion of a massive star due to an instability in its core. In section \ref{sec:mesamodels} we apply our derivations to models of supergiants which did not experience a rapid expansion. In section \ref{sec:sn2009ip} we discuss the application of our model for SN~2009ip, and some other similar transient events. We summarize our main findings in section \ref{sec:summary}.

\section{SCENARIOS}
\label{sec:scenarios}

The general scenario we consider is the passage of a NS through the envelope of a larger star, such as a supergiant. The NS is likely to be in an eccentric orbit, following the formation of the NS with a natal kick, and it plunges deep into the envelope only near periastron passages. Accretion onto the NS through an accretion disc is followed by an energetic bipolar outflow we term `jets', powering a luminous transient, or outburst. The three following scenarios can bring a NS into the envelope of a massive supergiant star.

(i) The supergiant experiences a phase of rapid expansion near the end of its evolution (e.g., \citealt{QuataertShiode2012,McleySoker2014,SokerGilkis2017a}). This might be observed as a \textit{pre-explosion outburst}, occurring just before the CCSN explosion of the supergiant.

(ii) The companion star reaches the end of its evolution and becomes a NS through a CCSN explosion, receiving a natal kick \citep{NSKicks1,NSKicks2,NSKicks3,NSKicks4} which brings its orbit to interact with the envelope of the supergiant, causing a \textit{post-explosion outburst}. This is conceptually similar to the scenario discussed by \cite{MichaelyPerets2018}, where a CCSN precedes the merger event of two compact objects.

(iii) A dynamical perturbation due to a tertiary star (e.g., \citealt{PeretsKratter2012}) changes the orbit of the inner binary, causing the NS to enter the envelope of the larger star. In this case the ensuing outburst might be unrelated directly to a CCSN explosion.

In the latter two scenarios, the envelope structure of the supergiant star will be similar. In principle, the engulfing star can also be a MS star, but we will focus on an evolved supergiant in this study. The envelope structure for the first scenario in the list above is expected to be different, as the envelope has expanded significantly due to energy deposition following an instability in the core.

In all the scenarios listed above, if the NS is captured in the envelope, it can spiral-in all the way to the core and completely disrupt the star (e.g., \citealt{SokerGilkis2018}). In this case an energetic terminal explosion occurs that is termed a CEJSN. Otherwise, the outburst is related or unrelated to a CCSN depending on which scenario, from those listed above, has brought the NS into the envelope of the supergiant.

A point which is relevant for all considered scenarios is the existence of a NS companion to a massive star, which can theoretically be more massive than the progenitor of the NS. This is possible if mass transfer occurred earlier in the evolution, with the initially more massive star transferring some material onto its companion before collapsing into a NS. Also, it might be that the relation between the initial mass and the compact remnant is non-trivial and non-monotonic. We will not discuss further this point.
 
A key assumption in our study is that a NS that accretes mass at a high rate and with sufficient angular momentum to form an accretion disc launches jets. There are several simulations that show the formation of an accretion disc around the compact object (a NS or a BH) that is formed from the collapsing core of a massive star (e.g., \citealt{MacFadyenWoosley1999, SekiguchiShibata2011, Tayloretal2011, BattaLee2016, Gilkis2018}). The two-dimensional simulations of \cite{Nagatakietal2007} which include magneto-hydrodynamic effects show the formation of jets with an energy of about $2 \times 10^{49} \erg$, supporting the notion that a NS or a BH can launch jets when accreting mass from a disc at a high rate and with sufficient angular momentum. \cite{Nagatakietal2007} conclude that their results cannot represent gamma-ray burst jets (see also \citealt{Fujimotoetal2006}), but this has no significance for our study since we do not look for jets with high Lorentz factors.
 
While the collapse of a massive star is simulated with high resolution by focusing on the inner core region, the accretion onto a compact object orbiting inside the envelope of its companion is considerably more difficult to fully simulate due to its inherent multi-scale nature. Yet we can find some inspiration from qualitatively similar scenarios. \cite{Staffetal2016} show in their hydrodynamical simulation of a MS companion in an eccentric orbit around an asymptotic giant branch star that near periastron an accretion disc is formed. A NS is much smaller than a MS star and much smaller than the resolution in their simulations. This implies that (i) even gas accreted with much less specific angular momentum can form an accretion disc, and (ii) the disc is tightly bound to the NS. The process is such that the accretion disc is formed before the NS enters the envelope, or when the NS is near the surface, and the disc survives as the NS orbits inside the envelope.

Further support comes from observations. \cite{BlackmanLucchini2014} suggest that the large momenta in some bipolar planetary nebulae (PNe) require that the energetic jets that inflated the lobes were launched inside a common envelope. In PNe the companion is most likely a MS star, or maybe a WD. It is easier still to form an accretion disc around a NS.

\section{The characteristics of the interaction}
\label{sec:parameters}

The proposed scenario is based on the possibility of a NS to accrete mass at very high rates thanks to cooling by neutrinos \citep{HouckChevalier1991, Chevalier1993, Chevalier2012}. Neutrino cooling is efficient when the mass accretion rate is $\dot M_{\rm acc} \ga 10^{-3} \rmModot \yr^{-1}$ \citep{HouckChevalier1991}. Furthermore, if jets are launched, as we assume in the present study, then cooling by jets takes away energy from the accretion disc.
 
To estimate the power of the jets, we assume that cooling by neutrinos plays the same role as cooling by photons (radiation) in traditional geometrically-thin accretion discs. In young stellar objects (YSOs), for example, the geometrically-thin accretion disc implies a very efficient radiative cooling. Despite this efficient cooling the canonical assumption for jets in YSOs is that they carry $\approx 10$--$40 \%$  of the accreted mass (e.g., \citealt{Federrathetal2014} and references therein), and their terminal velocity is about the escape speed from the YSO. We expect that the accretion disc in our studied case will be turbulent and contain strong magnetic fields much as geometrically-thin discs around YSOs. These ingredients are generally considered to be required for jet launching from accretion discs. In what follows, therefore, we assume that the jets carry a fraction of $\epsilon_j\approx 0.1$ of the accreted mass, and that their terminal velocity is about the escape velocity from a NS, $v_j \simeq 10^5 \km \s^{-1}$.
 
In the proposed scenario, a NS orbits a massive supergiant star. At a certain point, the NS finds itself inside the envelope, as discussed in section \ref{sec:scenarios}. The velocity of the NS relative to the envelope, $v_{\rm rel}$, will be about the Keplerian velocity. The mass accretion rate is estimated as    
\begin{equation}
\dot M_{\rm acc} \simeq \pi R^2_{\rm acc} \rho (r) v_{\rm rel}, 
\label{eq:dotMacc1}
\end{equation} 
where $\rho(r)$ is the envelope density at the location of the NS, and the accretion radius is given according to the Bondi paradigm as
\begin{equation}
R_{\rm acc} = \frac {2 G M_{\rm NS}}{v^2_{\rm rel} + c^2_s},  
\label{eq:Racc}
\end{equation} 
where $c_s(r)$ is the sound speed in the envelope and $M_{\rm NS}$ is the mass of the NS. As the envelope might rotate, the relative velocity is somewhat smaller than the orbital velocity of the NS. For the purpose of the present study we neglect the sound speed in equation (\ref{eq:Racc}); this increases the accretion radius. For estimating the relative velocity between the NS and the envelope we take the orbital velocity to be that of a circular orbit and neglect the rotation of the envelope. Doing so causes the accretion radius to decrease. Namely, these two assumptions contribute in opposite ways to the size of the accretion radius and about counterbalance each other, simplifying the calculation we perform. Substituting the orbital velocity for a massive star with mass $M_1 \gg M_{\rm NS}$ we derive a simple expression for the accretion rate. We scale the quantities with typical values, and derive 
\begin{eqnarray}
\begin{aligned}
\dot M_{\rm acc} & \simeq 0.18 
\left( \frac{M_{\rm NS}}{0.1 M_1} \right)^{2}
\left( \frac{r}{2 \AU} \right)^{2}  \\ 
&  \times 
\left( \frac{\rho(r)}{10^{-8} \g \cm^{-3}} \right)
\left( \frac{v_{\rm rel}}{100 \km \s^{-1}} \right)
\rmModot \yr^{-1} .
\label{eq:dotMacc2}
\end{aligned}
\end{eqnarray}
Over one orbit at this rate the NS accretes a mass of 
\begin{eqnarray}
\begin{aligned}
& M_{\rm acc} ({\rm orbit}) \simeq 8 \pi^2  
\left( \frac{M_{\rm NS}}{M_1} \right)^{2} r^3 \rho (r) \\
& = 0.11
\left( \frac{M_{\rm NS}}{0.1 M_1} \right)^{2}
\left( \frac{r}{2 \AU} \right)^{3}  
\left( \frac{\rho(r)}{10^{-8} \g \cm^{-3}} \right) \rmModot. 
\label{eq:MaccOrb}
\end{aligned}
\end{eqnarray}

We expect the envelope to have a shallow density profile when a pre-explosion star experiences a rapid expansion. For example, in the model that \cite{McleySoker2014} studied the density profile after the expansion can be approximated as 
\begin{equation}
\rho (r) \approx 3 \times 10^{-9} \left( \frac{r}{4 \AU} \right)^{-\beta} \g \cm^{-3}, 
\label{eq:rho}
\end{equation}
with $\beta \approx 1$. For a density profile with $\beta =1$ and for a NS reaching $r=0.7 R_1$, where $R_1$ is the radius of the supergiant, the envelope mass outside the radius $r$ is $M_{\rm e,out}\simeq 6.5 \rho (r) r^3$. As the NS orbits in the outer envelope, the jets that are launched by the NS will not interact directly with the envelope mass along the primary star's polar directions \citep*{Shiberetal2017}. The jets interact with a fraction $\epsilon_e$ of the envelope mass
\begin{equation}
M_{\rm e,int} \simeq 5 \epsilon_e \rho (r) r^3.
\label{eq:Meint}
\end{equation}
Note that the density profile as given by equation (\ref{eq:rho}) is for an inflated envelope caused by a short disturbance in the stellar interior. The density profile in the undisturbed envelope is much steeper (see section \ref{sec:mesamodels}).

We estimate the value of $\epsilon_e$ as follows. Since the NS moves through the envelope, the jets it launches do not punch a hole through the envelope and escape unimpeded. The jets are shocked and inflate large hot low-density bubbles, as seen in 3D hydrodynamical simulations of jets in common envelope evolution and of grazing envelope evolution (e.g., \citealt{Sokeretal2013, Shiberetal2017, LopezCamaraetal2018, ShiberSoker2018}). As the inflated bubbles make their way out of the envelope they interact with most of the envelope material in the region from the jets' origin (which changes its position) to the surface, including even some envelope gas near the equatorial plane. The condition for a jet to inflate a bubble rather than to readily escape is that its axis changes its location and/or direction on a timescale shorter than the time it takes for the jet to penetrate out from the envelope  (e.g., \citealt{Soker2016} for review). This condition is met here owing to the orbital motion of the NS. As the NS makes about half an orbit inside the giant, and because the bubble interacts with some material inward to its orbit, we approximately take $\epsilon_e \approx 0.5$.

Let a fraction $\epsilon_j$ of the accreted mass be launched in the jets at a velocity of $v_j$. Using equations (\ref{eq:MaccOrb}) and (\ref{eq:Meint}), we find the ratio of the mass in the jets to that in the envelope it interacts with to be 
\begin{eqnarray}
\begin{aligned}
\frac{M_j}{M_{\rm e,int}} & \simeq  
\frac {8 \pi^2 \epsilon_j   
\left( M_{\rm NS}/M_1 \right)^{2} } {5 \epsilon_e }\\
& \simeq 0.03 
\left( \frac{M_{\rm NS}}{0.1 M_1} \right)^{2}
\left( \frac{\epsilon_j}{0.1} \right)  
\left( \frac{\epsilon_e}{0.5} \right)^{-1} . 
\label{eq:MjMeint}
\end{aligned}
\end{eqnarray}
Conservation of energy implies that the jets eject the envelope with a typical velocity of 
\begin{eqnarray}
\begin{aligned}
v_{\rm e,ej} & \approx 
\left( \frac{M_j}{M_{\rm e,int}} \right)^{1/2} v_j 
\simeq 1.7 \times 10^4   
\left( \frac{M_{\rm NS}}{0.1 M_1} \right)
\\
& \times 
\left( \frac{\epsilon_j}{0.1} \right)^{1/2}  
\left( \frac{\epsilon_e}{0.5} \right)^{-1/2} 
\left( \frac{v_j}{10^5 \km \s^{-1}} \right) \km \s^{-1}. 
\label{eq:vjet}
\end{aligned}
\end{eqnarray}
This relation holds as long as the NS does not begin a second orbit and interacts again with the same envelope region as before, that is, for less than one orbit as occurs for example in a periastron passage. In the case of a periastron passage the NS crosses a shorter distance than a circumference, $\delta 2 \pi r$, with $\delta < 1$. The interaction lasts for a duration of 
\begin{eqnarray}
\begin{aligned}
 \tau_{\rm int} &\approx  \frac{\delta 2 \pi r_p}{v_{\rm rel}} 
= 2 \left( \frac{\delta}{0.3} \right) \\
& \times    
\left( \frac{r_p}{2 \AU} \right)   
\left( \frac{v_{\rm rel}}{100 \km \s^{-1}} \right)^{-1}
{\rm months},
\label{eq:tauint}
\end{aligned}
\end{eqnarray}
where $r_p$ is the orbital separation at periastron. The jets remove envelope mass from the vicinity of the NS and reduce the accretion rate, or even stop the accretion entirely. Namely, the interaction operates in a negative feedback mechanism. For that, the interaction time can be shorter than the value given by equation (\ref{eq:tauint}), and the typical ejected velocity somewhat smaller than that given by equation (\ref{eq:vjet}).

The energy carried by the jets is 
\begin{equation}
E\left(r\right) = \frac{1}{2} \epsilon_j \dot{M}_\mathrm{acc}\left(r\right) \tau_\mathrm{int} \left(r\right) v_j^2.
\label{eq:Emodels}
\end{equation}
Substituting equations (\ref{eq:MaccOrb}) and (\ref{eq:tauint}), we find for the total energy carried by the jets  
\begin{eqnarray}
\begin{aligned}
 E\left(r\right)  &\simeq 3 \times 10^{50}
\left( \frac{\delta}{0.3} \right)
\left( \frac{\epsilon_j}{0.1} \right) 
\\ & \times 
\left( \frac{v_j}{10^5 \km \s^{-1}} \right) ^2  
\left( \frac{M_{\rm NS}}{0.1 M_1} \right)^{2}
\\  & \times
\left( \frac{r_p}{2 \AU} \right)^{3}  
\left( \frac{\rho (r)}{10^{-8} \g \cm^{-3}} \right) \erg. 
\label{eq:Eoutburst}
\end{aligned}
\end{eqnarray}
This energy is the major contribution to the outburst energy, e.g., much larger than the binding energy of the envelope mass that is removed, and hence we consider it to be about the ILOT energy, $E_{\rm ILOT} \simeq E\left(r\right)$.

In our rudimentary derivations in this section we neglect the negative feedback mechanism through which the jets interact with the ambient medium. For that, we somewhat overestimate the interaction time and the outburst (ILOT) energy.

An important question is whether the estimated accretion rate (equation \ref{eq:dotMacc2}) is reasonable, and also how jets are launched from the disc. According to \cite{Chevalier1993}, the trapped radiation cannot stop the accretion rate from reaching the high rate required for neutrino cooling, and this is supported by the simulations of \cite{Fryeretal1996}. Furthermore, jets help the accretion by taking away high entropy material and angular momentum. The recent hydrodynamical common envelope simulations of \cite{Chamandyetal2018} show that if jets (as claimed by \citealt{Shiberetal2016}) or another process remove energy from near the mass-accreting star then accretion can proceed at very high rates, i.e., super-Eddington rates. Another issue is the requirement of the accretion shock to be smaller than the sonic radius of the Bondi accretion (see \citealt{BarkovKomissarov2011}), although we note that the shock radius can be smaller than analytic estimations, due to removal of angular momentum and high entropy gas by jets. Finally, we assume that neutrino cooling does not take away all of the accretion energy, as material outflow must remove angular momentum from the accretion disc. We assume that similar to other astrophysical objects accreting from a disc, the bipolar outflow will be at about the escape velocity and carry $\approx 10 \%$ of the accreted mass, i.e., $\epsilon_j \approx 0.1$, as implied in our scaling of equation (\ref{eq:Eoutburst}). The full details of accretion with neutrino cooling and jet launching will have to be studied in considerably arduous and advanced future simulations.

\section{Application to supergiant models}
\label{sec:mesamodels}

\begin{figure}
\includegraphics[width=0.5\textwidth]{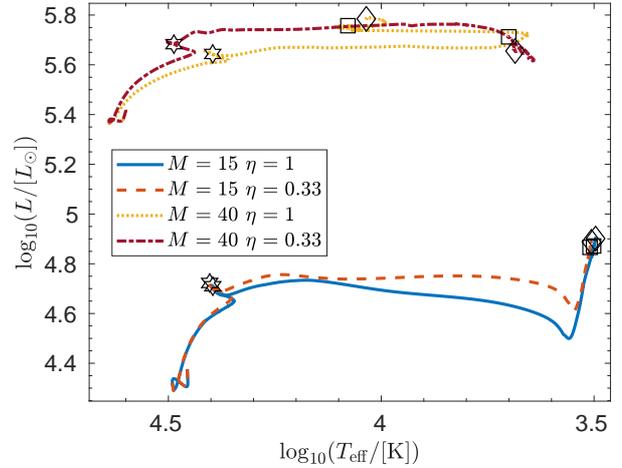}
\caption{Hertzsprung-Russell diagrams of the four supergiant star models we study. Hexagram symbols mark core hydrogen depletion, square symbols the depletion of helium in the core, and the depletion of carbon in the core is marked by diamond symbols.}
\label{fig:hr}
\end{figure}

To further demonstrate our proposed scenarios, we use values in the envelopes of stellar models evolved with Modules for Experiments in Stellar Astrophysics (\texttt{MESA} version 10108; \citealt{Paxtonetal2011,Paxtonetal2013,Paxtonetal2015,Paxtonetal2018}). The models are non-rotating and have a metallicity of $Z=0.02$. Mixing processes include convection according to the Mixing-Length Theory \citep{BohmVitense1958} with $\alpha_\mathrm{MLT}=1.5$, semiconvective mixing \citep{Langer1983, Langer1991} with $\alpha_\mathrm{sc}=0.1$, and exponential convective overshooting is applied as in \cite{Herwig2000} above and below non-burning and hydrogen-burning regions (with the fraction of the pressure scale height for the decay scale of $f=0.016$). We evolve two masses, $M_\mathrm{ZAMS}=15 \rmModot$ and $M_\mathrm{ZAMS}=40 \rmModot$, up to the stage of core carbon depletion. Mass loss is according to \cite{Vink2001} for the MS phase, and according to \cite{deJager1988} during the evolved supergiant phase, and we apply a multiplicative factor $\eta$ to the mass loss at all times (see, e.g., \citealt{Smith2014,Renzo2017}, on mass loss in massive stars). We use $\eta=0.33$ and $\eta=1$, for a total of four models, which we present in Fig. \ref{fig:hr}.

\begin{figure}
\includegraphics[width=0.5\textwidth]{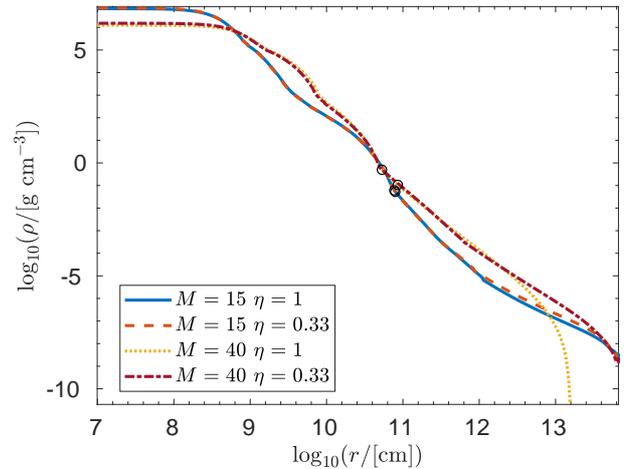}
\caption{Density profiles of the four models, at the stage of core carbon depletion. Black circles mark the transition from the core to the envelope, where the hydrogen fraction drops below $0.3$.}
\label{fig:profiles}
\end{figure}

In Fig. \ref{fig:profiles} we present the density profiles of the four models at the stage where carbon is depleted in the core. The two models with $M_\mathrm{ZAMS}=15 \rmModot$ are red supergiants (RSG) at this stage, while the models with $M_\mathrm{ZAMS}=40 \rmModot$ evolve into a yellow supergiant (YSG) for $\eta=0.33$, and a blue supergiant (BSG) for $\eta=1$. It can be seen in Fig. \ref{fig:profiles} that the BSG has a steeper density decline in the outer envelope compared to the other models.

To estimate the characteristics of an outburst powered by a NS interacting with the modeled envelopes, we proceed as follows. We start from equation (\ref{eq:Racc}) where we neglect the sound speed and substitute there and in equation (\ref{eq:dotMacc1}) ${v_\mathrm{rel}=\left(G m_r/r\right)^{1/2}}$ for the NS-envelope relative velocity, where $m_r$ is the mass of the supergiant inner to radius $r$. We derive the following equation for the accretion rate onto the NS  
\begin{equation}
\dot{M}_\mathrm{acc}\left(r\right) = 4 \pi \left(\frac{M_\mathrm{NS}}{m_r}\right)^2 r^2 \rho \left(r\right) \left(\frac{G m_r}{r}\right)^{1/2}.
\label{eq:mdotmodels}
\end{equation}
Substituting the relative velocity in equation (\ref{eq:tauint}) and taking for the reduced interaction time there $\delta\left(r\right)=1-r/R_1$, we find
\begin{equation}
\tau_\mathrm{int} \left(r\right) = 2 \pi \left( 1-\frac{r}{R_1} \right)
\left(\frac{r^3}{G m_r}\right)^{1/2}. 
\label{eq:tauintmodels}
\end{equation}

Taking for the mass in the jets $M_j=\epsilon_j \dot{M}_\mathrm{acc} \tau_\mathrm{int}$ and for the envelope mass that the jets interact with  $M_\mathrm{e,int}=\epsilon_e \left( M_1 - m_r \right)$, and substituting in equation (\ref{eq:vjet}), yields the following expression for the typical velocity of the ejected envelope 
\begin{equation}
v_\mathrm{e,ej} \left(r\right) \simeq \left[\frac{\epsilon_j \dot{M}_\mathrm{acc} \left(r\right) \tau_\mathrm{int}\left(r\right)}{\epsilon_e \left( M_1 - m_r \right)}\right]^{1/2} v_j.
\label{eq:veejmodels}
\end{equation}
Finally, for the outburst energy we use equation (\ref{eq:Emodels}).  

\begin{figure}
\includegraphics[width=0.5\textwidth]{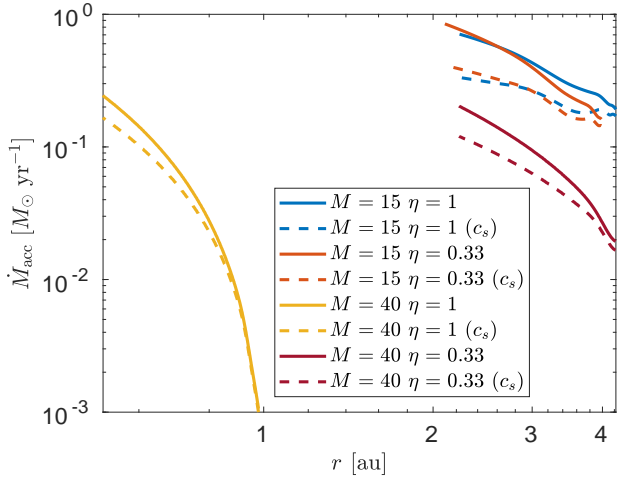}
\caption{The accretion rate calculated by equation (\ref{eq:mdotmodels}) for four different stellar models (solid lines), for a NS moving inside the envelope at an orbital separation of $0.5 R_1< r <0.95 R_1$. We also present by the dashed lines the accretion rates when the sound speed is considered in the expression for the accretion radius (see text).}
\label{fig:accretion}
\end{figure}

We first apply equation (\ref{eq:mdotmodels}) to our stellar models, and present the results in Fig. \ref{fig:accretion}. We note that the passage of the NS through the envelope will not be in a circular trajectory, and therefore not at constant $r$. However, as we see in Fig. \ref{fig:accretion}, the accretion rate is not very sensitive to the depth within the envelope from which we take the values for equation (\ref{eq:mdotmodels}). We see that $\dot{M}_\mathrm{acc}\left(r\right)>10^{-3} \rmModot \yr^{-1}$ for essentially all values of $r$ in all models, as required for efficient cooling by neutrinos.
  
We also checked the effect of not neglecting the sound speed $c_s$ in the expression for the accretion radius (equation \ref{eq:Racc}) that we used in the derivation of equation (\ref{eq:mdotmodels}). As shown in Fig. \ref{fig:accretion}, this has a limited effect. Furthermore, the uncertainty in the relative velocity due to the rotation of the supergiant might bring the accretion rate back up to around the values calculated without taking $c_s$ in the derivation.

\begin{figure}
\includegraphics[width=0.5\textwidth]{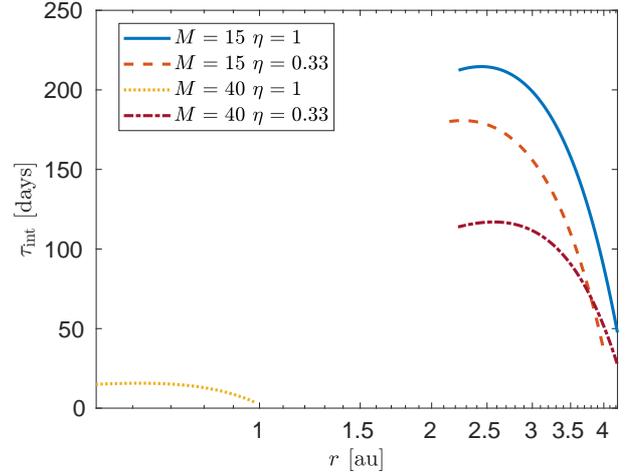}
\caption{The interaction time according to equation (\ref{eq:tauintmodels}) for our four different stellar models and for periastron orbital separation of $0.5 R_1 < r <0.95 R_1$.}
\label{fig:interactiontime}
\end{figure}

In Fig. \ref{fig:interactiontime} we show the duration of the interaction of the NS with the supergiant envelope, calculated using equation (\ref{eq:tauintmodels}). For the models and parameters we employ the interaction times range from days to several months. For the RSG and YSG models the interaction time when the NS does not get deep into the envelope is about a month. The interaction time can last for about half a year when the NS dives deep into the envelope. For the BSG model the interaction lasts between only days to a few weeks, due to its smaller size. As mentioned in section \ref{sec:parameters}, the duration might be overestimated in all cases due the feedback nature of the interaction. 

\begin{figure}
\includegraphics[trim= 0.3cm 1.1cm 1.1cm 1.1cm,clip=true,width=1.0\columnwidth]{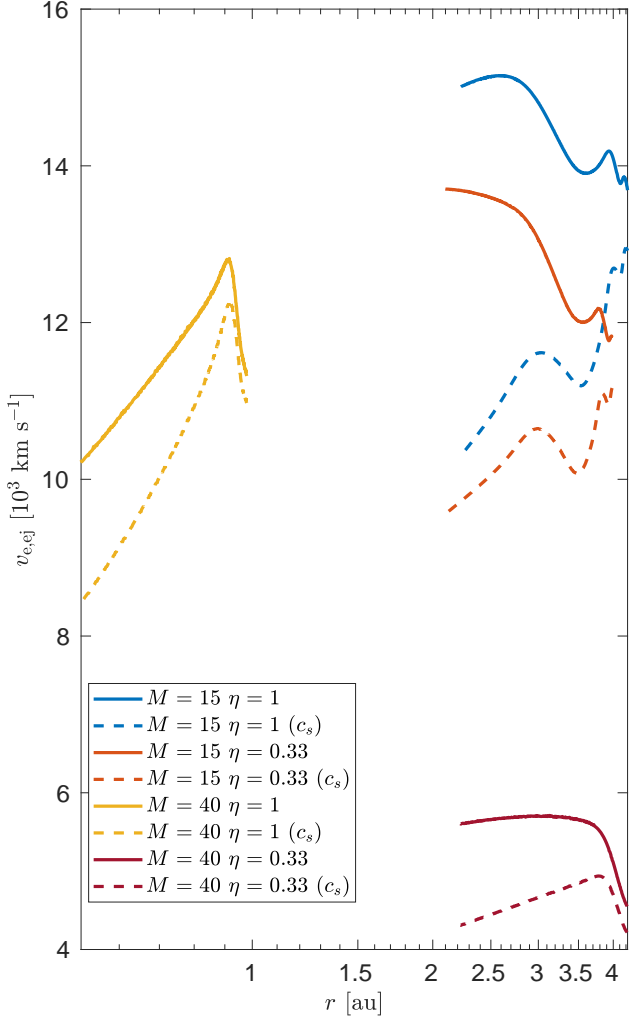}
\caption{The ejecta velocity calculated for four different stellar models, and using equation (\ref{eq:veejmodels}) with $\epsilon_j=0.1$, $\epsilon_e=0.5$ and $v_j=10^5 \kms$ (solid lines), as function of the orbital separation in the range of $0.5 R_1 < r <0.95 R_1$. The effect of taking into account also the sound speed is shown in the dashed lines.}
\label{fig:vejecta}
\end{figure}

In Fig. \ref{fig:vejecta} we show the estimated velocity of the ejecta from the interaction, using equation (\ref{eq:veejmodels}), with $\epsilon_j=0.1$, $\epsilon_e=0.5$ and $v_j=10^5 \kms$. The range of velocities is between $4\,000 \kms$ and $16\,000 \kms$, differing between stellar models. The sensitivity to $r$ in each model is not large. Taking into account the sound speed $c_s$ in the expression for the accretion radius changes the velocities somewhat. We expect realistic values to be between those calculated with and without the inclusion of $c_s$. 
    
\begin{figure}
\includegraphics[width=0.5\textwidth]{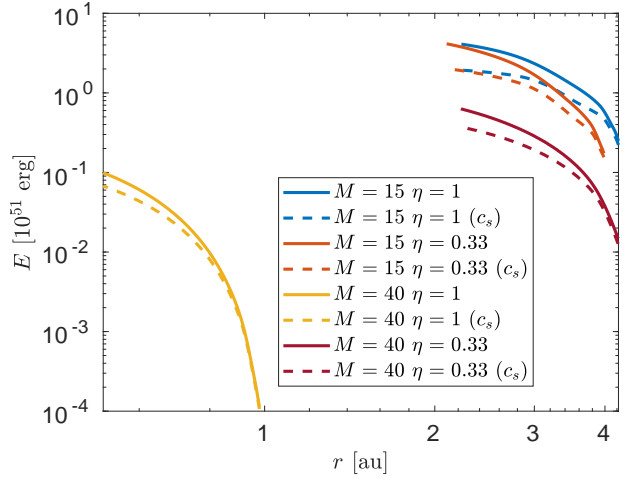}
\caption{The outburst energy according to equation (\ref{eq:Emodels}) with $\epsilon_j=0.1$ and $v_j=10^5 \kms$ (solid lines), for our four stellar models, and in the range $0.5 R_1 < r <0.95 R_1$. The effect of taking into account also the sound speed is shown in the dashed lines.}
\label{fig:energy}
\end{figure}

In Fig. \ref{fig:energy} we show the outburst energy estimated using equation (\ref{eq:Emodels}), with $\epsilon_j=0.1$ and $v_j=10^5 \kms$. Similar to our estimation of the interaction duration (Fig. \ref{fig:interactiontime}), we somewhat overestimate the outburst energy due to a negative feedback mechanism through which the jets interact with the ambient gas \citep{Soker2016}. The very high values of $E>10^{51} \erg$ are therefore not realistic. Outburst energies of a few times $10^{50}$, though, are reasonable.

The results we show in this section are for the stage at which carbon is depleted in the core, just several years before the final collapse of the iron core. We also checked an earlier stage, that of core helium depletion, which is several thousands of years earlier for the $M_\mathrm{ZAMS}=40 \rmModot$ models, and a few tens of thousands of years earlier for the $M_\mathrm{ZAMS}=15 \rmModot$ models. We found  very small quantitative differences in $\dot{M}_\mathrm{acc}$, $\tau_\mathrm{int}$, $v_\mathrm{e,ej}$ and $E$. Therefore, our results are not sensitive to the precise evolutionary stage of the supergiant.

The passage of the NS through the envelope of the giant will result in changes to the orbital parameters, with consequences also for subsequent such passages. For the purpose of demonstration, we take a typical accretion rate of ${ \dot{M}_\mathrm{acc} \simeq 0.3 \rmModot \yr^{-1} }$ (see Fig. \ref{fig:accretion}) and an interaction time of $\tau_\mathrm{int} \simeq 0.5 \yr$ (see Fig. \ref{fig:interactiontime}), and consider a plunge to a depth of $r \simeq 3 \AU$. The accreted mass is then ${ M_\mathrm{acc} \simeq 0.15 \rmModot \simeq 0.1 M_{\rm NS} }$, similar to the scaled accretion mass in equation (\ref{eq:MaccOrb}). From angular momentum conservation an accretion of such a mass (with zero angular momentum) acts to reduce the semi-major axis of the orbit by about $20 \%$. Since we expect the envelope to rotate in the same direction as the orbital motion of the secondary star, the accreted gas has some angular momentum. This reduces the effect of the accretion on the orbit. Further offsetting the decrease of the semi-major axis, the jets remove a gas mass of about 30 times their own mass from the envelope (equation \ref{eq:MjMeint}), or about 3 times the accreted mass. Overall we estimate the orbit to shrink by about $10 \%$ in one passage.

Another effect which might change the orbit of the NS is that of dynamical friction. We estimate the gravitational drag force (see, e.g., \citealt{Ostriker1999}) as ${ F_\mathrm{dyn} \approx 4 \pi G^2 M_\mathrm{NS}^2 \rho / v_\mathrm{rel}^2 }$, and then the relative velocity change ${ \delta v / v \approx (F_\mathrm{dyn} / M_\mathrm{NS}) \tau_\mathrm{int} \left( r \right) / v_\mathrm{rel} }$ is about a few percent for most of the parameter range considered, and up to $ \approx 20 \% $ for the deepest plunges. This is similar to the kinematic effect of the mass transfer.

The effects described above cause a moderate decrease in the orbital separation, so that the NS can survive several such passages inside the envelope, with some sensitivity to uncertainties in the binary evolution. For example, the basic expectation is that the time between successive outbursts will decrease due to accretion. However, if we consider that the interaction might have been triggered by some instability that caused the inflation of the envelope, then the binding energy of the disturbed envelope might be very low. Both the jets launched by the NS inside the envelope, and the stellar wind after the NS exits the envelope following its eccentric orbit, might remove large amounts of mass such that the orbital period might even increase. This is similar to the case of the orbital separation increasing in the grazing envelope evolution \citep{Soker2017}.  

\section{SN 2009\lowercase{ip}}
\label{sec:sn2009ip}
  
\subsection{Observational properties}

The supernova impostor SN~2009ip first erupted in 2009, soon to be discovered as an impostor of LBV origin, rather than a supernova \citep{Maza2009,Berger2009}. The LBV, located in the spiral galaxy NGC~7259, showed a series of outbursts, the first of which in 2009 and the last in 2012 (e.g., \citealt{Drake2012,Mauerhanetal2013,Pastorello2013,Levesqueetal2014}). The outbursts showed an increase by $\approx 3$--$4$ magnitudes in the V~band in Sep. 2011 and Aug. 2012 (hereafter outburst 2012a), followed by an increase of $\approx 7$ magnitudes in Sep. 2012 (hereafter outburst 2012b). The peak bolometric luminosity of the 2012b outburst at initially estimated to be $L_{\rm p} = 8 \times 10^{42} \erg \s^{-1}$ \citep{Pastorello2013}. 

Assuming the erupting star was a non-rotating LBV, \cite{Foley2011} suggested that the ZAMS mass of the erupting star was $M_1 \geq 60 \rmModot$. Assuming a rotating LBV at $40\%$ of its critical velocity, \cite{Marguttietal2014} gave an estimate of $M_1 = 45$--$85 \rmModot$. A later estimate based on multi-spectral observations of the outburst updated the value to $L_{\rm p} = 1.2 \times 10^{43} \erg \s^{-1}$ \citep{Marguttietal2014}. Consequently, the bolometric energy radiated during the outbursts was found to be $E_{{\rm rad,}a}=(1.5 \pm 0.4) \times 10^{48} \erg$ for the 2012a outburst, and $E_{{\rm rad,}b}=(3.2 \pm 0.3) \times 10^{49} \erg$ for the 2012b outburst \citep{Fraseretal2013,Marguttietal2014}. The total energy involved in each of the outbursts is a few times larger than this value and was estimated to be $E_{{\rm tot,}a}=(2 \pm 1) \times 10^{48} \erg$ and $E_{{\rm tot,}b}=(7.5 \pm 2.5) \times 10^{49} \erg$ \citep*{Kashietal2013, Marguttietal2014}.

\cite{Marguttietal2014} suggested that most of the energy radiated in the large 2012b peak came from the kinetic energy of the material ejected during the 2012a outburst. Calibrating the ejecta mass with $\approx 0.5 \rmModot$, \cite{Marguttietal2014} found the total energy of the outbursts to be $E_{{\rm tot}}\approx 10^{50} \erg$. An important characteristic of SN~2009ip which is relevant to our present study is the high ejecta velocity of the 2011 eruption (up to $\approx 13\,000 \kms$; \citealt{Pastorello2013}) and of the 2012a event \citep{SmithMauerhan2012, Mauerhanetal2013}. 

\cite{Mauerhanetal2014} observed SN~2009ip during the 2012a outburst and found that the the spectrum showed broad P~Cyg lines. They found polarization that suggests substantial asphericity for the 2012a outflow. The degree of polarization increased during the 2012b event, from which \cite{Mauerhanetal2014} concluded a higher degree of asphericity than 2012a. The asymmetry was later confirmed by observations of \cite{Reillyetal2017}.

\cite{Fraseretal2015} followed the decline of the light curve in 2013--2014, and found that its slope was considerably shallower than expected from nuclear decay slopes of CCSNe. From the spectroscopic and photometric evolution until 820 days after the initiation of the 2012a event, they found no evidence that a CCSN had occurred. \cite{Grahametal2014} and \cite{Grahametal2017} also presented observations of the late evolution of the light curve (the later up to 1000 days post-eruption). They found that the light curve is still decreasing in a linear rate, an expected behavior for eruptions interacting with circumstellar material (CSM). They also compared late spectra of SN~2009ip to various SNe and SN impostors. They could not conclusively tell whether the interaction with the material is the result of an impostor or a real supernova, but found it somewhat better matches a real supernova.

\subsection{Previously proposed models}

\cite{Ouyedetal2013} attributed the 2012a outburst to a standard CCSN, and the 2012b outburst to a dual-shock quark-nova. \cite{Mauerhanetal2013} proposed a second scenario, suggesting that the 2012a event was a terminal supernova explosion, and the 2012b outburst to be the result of collision of fast supernova ejecta from the 2012a outburst with slower gas ejected earlier. The same scenario was also favored by \cite{Prieto2013}. \cite{Marguttietal2014} attributed the 2012b brightening to an explosive shock breakout coming from an interaction between the explosive ejection of the LBV envelope taking place $\approx 20$--$24$ days before the 2012b peak, and shells of material ejected during the 2012a eruption. The results of \cite{Marguttietal2014} disqualify the \cite{Mauerhanetal2013} scenario. The reason, as noted by \cite{Marguttietal2014}, is that the photosphere expansion velocity of $\approx 4500 \km \s^{-1}$ during the 2012b outburst implies that the gas that accelerated the photosphere originated long after the peak of the 2012a event. Namely, the gas was ejected long after the star had ceased to exist according to \cite{Mauerhanetal2013}.

Another scenario favored core instability of a single star that leads to the ejection of shells \citep{Pastorello2013}. A different scenario was suggested by \cite{SokerKashi2013}, who compared the 2012a and 2012b outbursts to the outburst of the ILOT V838~Mon. The latter ILOT is composed of three shell-ejection episodes as a result of a stellar merger event \citep{Tylenda2005}. The ejection of separate shells in the 2012a and 2012b outbursts supports the binary scenario proposed by \cite{SokerKashi2013}, who suggested that SN~2009ip was a massive binary system with an LBV of $M_1=60$--$100 \rmModot$ and a MS companion of $M_2=0.2$--$0.5 M_1$ in an eccentric orbit.

\cite{Kashietal2013} proposed that the major 2012 outburst was powered by an extended and repeated interaction between the LBV and a more compact (MS or Wolf-Rayet star) companion in an eccentric orbit. During the first periastron passage, the companion accreted $2$--$5 \rmModot$ from the LBV envelope. The accreted gas released gravitational energy which can account for the total 2012b outburst energy. Also, in the declining light curve of the 2012b outburst \cite{Kashietal2013} noticed two large peaks in which the extra radiation was similar to the 2009--2011 outbursts. \cite{Kashietal2013} interpreted the peaks as resulting from mass ejected during later periastron passages. In that case the inferred orbital period after the large mass accretion is $\approx 25$ days, suggesting that the companion survived the eruption. 

In an additional scenario, \cite{Kashietal2013} considered a terminal binary merger event, but one which occurred only after the system had experienced a second periastron passage after the major one. As in the surviving companion scenario, the major interaction that powered the 2012b outburst was powered by mass accretion, which shortened the orbital period. However, in the merger scenario the orbit was shortened even more, and the second periastron passage occurred $\approx 20 \days$ after the first (major) periastron passage. After the second periastron passage the companion plunged too deep into the envelope to eject more gas.

\cite{Levesqueetal2014} found evidence for the existence of a thin disc around the central star, and suggested that a binary companion is also present. They favored a model in which the observed 2012b re-brightening is an illumination of the disc's inner rim by fast-moving ejecta produced by the underlying events of 2012a.

One of the challenges for the binary model is to account for gas moving at  $v>10\,000 \kms$ as observed in the 2011 outburst \citep{Pastorello2013} and in the 2012a outburst \citep{Mauerhanetal2013}. \cite{TsebrenkoSoker2013} simulated part of the scenario of \cite{Kashietal2013}, in which jets that are launched by the accreting companion and interact with the environment account for the high velocity gas. Namely, they numerically studied the propagation of the jets through the extended envelope. They were able to reach the observed velocities but only with a small fraction of the gas, probably smaller than can account for the observations. They also commented that jets launched by a WR companion will be narrower and denser than by a MS star, with a shorter flow time and a longer photon diffusion time, which would allow the acceleration of more mass to higher velocities.

\subsection{Applying our scenario for SN~2009ip}

We examine whether our proposed common envelope jets supernova impostor scenario can account for the observations of SN~2009ip. Namely, we examine the possibility that the 2011 outburst, and possibly the 2012 eruption, of SN~2009ip were supernova impostors driven by jets from a NS companion. The main advantage of jets from a NS companion is that they can account for the velocity of about $13,000 \km \s^{-1}$ that \cite{Pastorello2013} found in the 2011 outbursts. We will here refer to the 2012a event of SN~2009ip as the supernova, and will adopt the idea that the 2012b event is the result of an interaction with the CSM, ejected earlier.

We evolve a \texttt{MESA} model for an LBV starting from $M_{\rm ZAMS} = 110 \rmModot$, and having a mass-loss rate according to the prescription of \cite{Kashietal2016}, that keeps the photosphere temperature at $20\,000 \K$, in accordance with the bi-stability jump. We evolve it until it reaches $M_{\rm LBV} \simeq 80 \rmModot$. As LBVs are hot stars, their typical radius is smaller than that of RSGs. Therefore, the scenario proposed here requires the NS companion to be closer to the LBV than it would have been had the primary star been a RSG.

For our scenario we adopt the parameters developed by \cite{Kashietal2013}. As mentioned above, the mass of the LBV is $M_{\rm LBV} \simeq 80 \rmModot$, the orbital period is taken to be $P \approx 32$ days (note that the number stated earlier, $\approx 25$ days, is the period at the end of the interaction rather than the period before/during the interaction) and the interaction time is $\tau_{\rm int} \approx 8$ days, therefore $\delta \simeq 0.25$ (see equation \ref{eq:tauint}). The eccentricity is taken to be $e \simeq 0.5$ so that the NS reaches a periastron distance of $\approx 0.4 \AU$, that is inside the envelope of the LBV whose radius is $R_{\rm LBV} \simeq 0.55 \AU$. The mass of the NS companion is $M_\mathrm{NS} = 1.33 \rmModot$.

We use equation (\ref{eq:Eoutburst}) to calculate the total energy that the jets carry 
\begin{equation}
\begin{split}
E_j &\simeq  1.7 \times 10^{48}
\left( \frac{\delta}{0.25} \right)
\left( \frac{\epsilon_j}{0.1} \right) \\ 
& \times 
\left( \frac{v_j}{10^5 \km \s^{-1}} \right) ^2  
\left( \frac{60 M_{\rm NS}}{ M_{\rm LBV}} \right)^{2}\\  
& \times
\left( \frac{r}{0.4 \AU} \right)^{3}  
\left( \frac{\rho (r)}{3 \times 10^{-7} \g \cm^{-3}} \right) \erg. 
\end{split}
\label{eq:Ej_SN2009ip}
\end{equation}
This energy is about the energy released in the 2012a event. We conclude that even with conservative parameters the accretion onto a NS from the LBV envelope can account for the observed energy in SN~2009ip and probably also other type~IIn supernovae.

The series of six peaks observed in 2011 (that should have probably had seven peaks as one was evidently missing) had intervals of $\approx 40$ days. It is hard to tell what the duration of interaction was as observations are not frequent enough, but $8$ days is a reasonable duration to assume (note that \citealt{TsebrenkoSoker2013} considered a shorter interaction of 6--12 hours for each peak, which would yield 0.25--0.5 for the value of $\delta$ we use here for the entire episode of seven peaks in 2011). Therefore the same scaling of equation (\ref{eq:Ej_SN2009ip}) can also apply to the series of eruptions in 2011, which also had the same (combined) energy as the 2012a event.

Mass removal will also affect the orbit. For the present consideration, the NS enters the envelope of an LBV star, and not an inflated envelope as the one studied in section \ref{sec:parameters}, and hence equation (\ref{eq:Meint}) does not apply. The density profile of the envelope is steep, with $\beta \ga 3$, and for the model we use $\rho(r) \simeq 3 \times 10^{-7} (r/0.4 \AU)^{-\beta}$ (compare to  equation \ref{eq:rho}). This implies that the envelope mass outside a radius which equals the periastron distance of $0.4 \AU$ is $\approx 2 M_\odot$. The NS crosses about quarter of a circle, i.e., $\delta \simeq 0.25$, and hence the jets and the bubbles they inflate interact with $M_{\rm e,int} \approx 0.5 M_\odot$. Not all this mass is ejected as the jets marginally have the required energy to eject such a mass. More reasonably a mass of $\Delta M_{\rm ej} \approx 0.1$--$0.3 M_\odot$ is ejected at each periastron passage. After seven periastron passages the NS is expected to eject $\approx 1$--$2 M_\odot$. Mass removal at periastron passages increases the eccentricity, but as here the NS with its jets and the bubbles they inflate are expected to remove only $1$--$2 \%$ of the primary mass, the effect of mass loss on eccentricity is low.

In order to determine whether or not drag forces may cause the orbit to become circularized, we estimate the circularization timescale, given by \cite{VerbuntPhinney1995} as
\begin{equation}
\begin{split}
\frac{1}{\tau_c} \equiv -\frac{d\ln{e}}{dt}
&= 12.4 \left( \frac{T_{\rm eff}}{20\,000 \K} \right)^{4/3}
\left( \frac{M_{\rm env}}{\rmModot} \right)^{2/3}   \\
&\times \frac{\rmModot}{M_{\rm LBV}} \frac{M_{\rm NS}}{M_{\rm LBV}}
\frac{M_{\rm LBV}+M_{\rm NS}}{M_{\rm LBV}}
\left( \frac{R_{\rm LBV}}{a} \right)^{8} \rm{yr^{-1}}  ,    \\
\end{split}
\label{eq:tau_c}
\end{equation}
where $M_{\rm env} \simeq 0.9 M_{\rm LBV}$ is the mass of the LBV's envelope and $T_{\rm eff}$ is its effective temperature. We get ${\tau_c} \simeq 440 \yr$, and conclude that tidal circularization is insignificant on the timescale of the six or seven eruptions of SN~2009ip in 2011.

When the NS is inside the envelope, the tidal interaction considered in equation (\ref{eq:tau_c}) is for the mass inner to the location of the NS. Since the core of an LBV is much denser than its envelope $M_{\rm LBV}$ can stay as is, but the envelope mass should be replaced by its fraction inner to the NS position, $M_{\rm env,in}$. Two other effects that contribute to the drag on the NS then become important, the mass accretion onto the NS and the local gravitational interaction within the envelope mass that is not accreted. Both of these effects are of the same order of magnitude (section \ref{sec:mesamodels}). The fractional change in the eccentricity depends on the exact amount of mass that is accreted at each time along the orbit. For accretion near periastron, the decrease in eccentricity is of the order of the accreted mass divided by the NS mass, $\Delta e_{\rm acc} \approx - 2 M_{\rm acc} / M_{\rm NS}$. For the typical values used in equation (\ref{eq:Ej_SN2009ip}), we find the accreted mass to be $M_{\rm acc} \simeq 2 \times 10^{-4} M_\odot$. Namely, the change of eccentricity per orbit is $\Delta e_{\rm acc} \approx - 2 \times 10^{-4}$. Taking twice as large an effect due to gravitational interaction with mass that is not accreted, we find that for an orbital period of $P \approx 32$~days the timescale to change the eccentricity is $\tau_{c-{\rm drag}} \equiv \vert (\dot e_{\rm drag})^{-1} \vert \approx 150 \yr$. This is similar to the tidal timescale inside the envelope for our parameters (equation \ref{eq:tau_c}). Adding all these effects together we find that the typical time for circularization is $\tau_{c-{\rm tot}} \approx 100 \yr$.

The drag and mass accretion act to decrease the eccentricity. Enhanced mass loss at periastron passages, on the other hand, increases the eccentricity. As discussed above, at each periastron passage a mass of $\Delta M_{\rm ej} \approx 0.1$--$0.3 M_\odot$ is ejected from the envelope. This increases the eccentricity by $\Delta e_{\rm ej} \approx \Delta M_{\rm ej} / M_{\rm LBV} \approx 0.003$. However, the mass ejection lasts for a longer time than accretion and the effect decreases as the NS moves away from periastron. Overall, for our parameters, the mass-loss rate increases the eccentricity by $\Delta e_{\rm ej} \approx 10^{-3}$ each periastron passage (assuming the eccentricity is small), which for an orbital period of $P \approx 32$~days gives a decircularization timescale of $\tau_{\rm dc} \approx 100 \yr$. As was suggested for other cases (e.g., by \citealt{KashiSoker2018}, for the binary system HD~44179), the enhanced mass-loss rate induced by jets more or less compensates for the effect of the forces acting to decrease the eccentricity near periastron passages.

We conclude that the process we propose here of a NS launching jets can account for the occurrence of pre-explosion outbursts in SN~2009ip and similar objects, including those with outflow velocities of $\ga 10^4 \kms$.

\section{SUMMARY AND DISCUSSION}
\label{sec:summary}
 
In many scenarios for the formation of binary NS systems that eventually merge, the system experiences an early common envelope phase of a NS inside the envelope of a giant (e.g., \citealt{Chruslinskaetal2018}). We set the goal to examine the possible observational signatures of this phase. When a full common envelope phase takes place and the NS spirals in all the way to the core, the outcome might be a terminal supernova-like event \citep{Chevalier2012}, that \cite{SokerGilkis2018} termed a common envelope jets supernova (CEJSN). \cite{SokerGilkis2018} suggested that the peculiar supernova iPTF14hls was a CEJSN event. In the present study we considered cases where the NS can enter the envelope and then exit, so the outburst might repeat itself.
 
Essentially, the process is like that of many other ILOTS where a companion star accretes mass through an accretion disc and launches jets. The radiation comes directly from the accretion process, or, more likely, from the interaction of the jets with the ambient gas. In most cases of this high-accretion-powered ILOT (HAPI) model the companion was taken to be a MS (or slightly evolved) star \citep{KashiSoker2016,SokerKashi2016}. The new addition described in the present paper is the consideration of a NS companion. 

A NS companion introduces three essential differences from a MS companion: (i) When the accretion rate is above about $10^{-3} \msyr$, neutrino cooling allows accretion much above the Eddington accretion rate \citep{HouckChevalier1991}. Cooling by jets carries away more energy from the accretion disc and further eases the accretion. This implies that the outburst can be very energetic, up to supernova energies. For that reason we term this event a CEJSN impostor. (ii) The high velocity jets imply that in some cases outflow velocities of the ejecta above about $10^4 \km \s^{-1}$ might be observed (equation \ref{eq:vjet}). (iii) The mass of the NS is generally smaller than that of the MS companion in the HAPI model of LBV ILOTs, like Eta Carinae. This implies that even if the NS that is on an eccentric orbit and due to its high velocity when plunging into the envelope it is able to survive several orbits, it will eventually enter the envelope and perform the inevitable full common envelope evolution. In this case the energy of the outburst will be larger, and the event will be a CEJSN. 
 
Let us elaborate on the last point. In section \ref{sec:scenarios} we discussed several scenarios for the NS to enter the envelope. There are two possible cases for the primary giant star to find itself far from its terminal nuclear evolution. In the first case the NS was formed in a CCSN and the natal kick sent it into the envelope of the giant, and in the second case a the system experienced a perturbation by a tertiary star. In both cases the orbit became eccentric. After one or more periastron passages the NS can either remove the envelope and end in a tight orbit around the core that will later form another NS, or it can spiral into the core and lead to a very energetic CEJSN. \cite{Papishetal2015} raised the possibility that strong \textit{r}-process nucleosynthesis (that form the third peak of the \textit{r}-process) takes place in jets launched in such circumstances (see also \citealt{SokerGilkis2017b}). In cases where the primary giant star is about to explode, the CEJSN impostor will be followed by a CCSN.

In section \ref{sec:parameters} we derived scaled relations to show the typical expected properties for our CEJSN impostor scenario and their dependencies on different parameters, with the focus on stars which undergo rapid expansion near the end of their nuclear evolution. In section \ref{sec:mesamodels} we applied our scenario for envelopes of evolved supergiant star models, and found that the accretion rate is in the range where neutrino cooling is efficient ($\dot{M}_\mathrm{acc} > 10^{-3} \msyr$), as well as a limited sensitivity to the depth in the envelope where the NS passes. We found ejecta velocities between $4\,000 \kms$ and $16\,000 \kms$, interaction duration times from days to months, and output energies up to about $10^{51} \erg$.

In section \ref{sec:sn2009ip} we discussed SN~2009ip that had several LBV outbursts (supernova impostors) before its terminal explosion. We raised the possibility that the ejecta velocity of $> 10^4 \km \s^{-1}$ in the 2011 and 2012a outbursts were derived by jets from a NS companion. Though it is not possible to conclusively tell whether the terminal explosion (either 2012a or 2012b) was a CCSN or a spiraling of the NS toward the core (i.e., a CEJSN), our analysis shows that the energy released in the 2012a event is of the order of what would be expected from the scenario.

We call for a serious consideration of peculiar supernovae and impostors as outcomes of CEJSNe (energetic and terminal) and CEJSN impostors (that might repeat). With the operation of jets that are launched by a more compact companion, here a NS, that accretes mass from a giant we connect these types of outbursts to other ILOTs that are driven by accreting MS stars.

\section*{Acknowledgments}

We thank an anonymous referee for suggestions that substantially improved the presentation of our proposed scenario. This research was supported by the Asher Fund for Space Research at the Technion, and the Israel Science Foundation. AG gratefully acknowledges the support of the Blavatnik Family Foundation. AK thanks the support of the Authority for Research \& Development in Ariel University and the Rector of Ariel University.

\bibliographystyle{mnras}

\label{lastpage}
\end{document}